\documentclass[%
 reprint,
 amsmath,
 amssymb,
 amsfonts,
 aps,
 longbibliography,
]{revtex4-2}

\usepackage[dvipdfmx]{graphicx}
\usepackage{dcolumn}% Align table columns on decimal point
\usepackage{bm}% bold math
\usepackage{color}
\usepackage{algorithm2e}
\usepackage{tabularx}
\usepackage{multirow}
\usepackage{url}

\def\mat#1{#1}

\def\ket#1{\mbox{\boldmath $#1$}}

\newcommand{\MLE}{maximum-likelihood estimate}
\newcommand{\MLA}{minimum linear arrangement}
\newcommand{\ORGM}{ORGM}
\newcommand{\SBM}{SBM}
\newcommand{\RCM}{reverse Cuthill-McKee algorithm}

\begin{document}

%\preprint{}

\title{Finding community structure using the ordered random graph model}


\author{Masaki Ochi}
\email{ochi@iis.u-tokyo.ac.jp}
 \affiliation{Department of  Physics, The  University of Tokyo,\\
 5-1-5 Kashiwanoha, Kashiwa, Chiba, 277-8574, Japan}
 
\author{Tatsuro Kawamoto}%
 \email{kawamoto.tatsuro@aist.go.jp}
\affiliation{Artificial Intelligence Research Center, \\
  National Institute of Advanced Industrial Science and Technology, 2-3-26 Aomi, Koto-ku, 
  Tokyo, 135-0064, Japan }

\date{\today}

\begin{abstract}
Visualization of the adjacency matrix enables us to capture macroscopic features of a network when the matrix elements are aligned properly.
Community structure, a network consisting of several densely connected components, is a particularly important feature, and the structure can be identified through the adjacency matrix when it is close to a block-diagonal form. 
However, classical ordering algorithms for matrices fail to align matrix elements such that the community structure is visible.
In this study, we propose an ordering algorithm based on the maximum-likelihood estimate of the ordered random graph model.
We show that the proposed method allows us to more clearly identify community structures than the existing ordering algorithms.
\end{abstract}

%\keywords{}

\maketitle

\section{Introduction \label{sec:Intro}}

In a structural analysis of networks, extracting macroscopic structures from network data is a central task.
For instance, there is a plethora of studies on extracting a community structure, which is commonly defined as a network that consists of several sets of densely connected vertices \cite{girvan2002community,rosvall2008maps,fortunato2010community,estrada2012physics,peixoto2014hierarchical,fortunato2016community,fortunato202220}. 
A banded structure is another type of macroscopic structure, in which the vertices with close indices are more likely to be connected. 
Extraction of the banded structure is referred to as the linear arrangement or graph layout problem in graph theory \cite{Harper1964,Juvan1992,Diaz2002,Petit2003,Rao2005,Devanur2006,Caprara2011,seitz2010contributions}, also termed envelope reduction problem \cite{Juvan1992,Barnard95,George1997,DingHe2004}. 

\begin{figure}[t!]
  \centering
  \includegraphics[width=0.8\columnwidth]{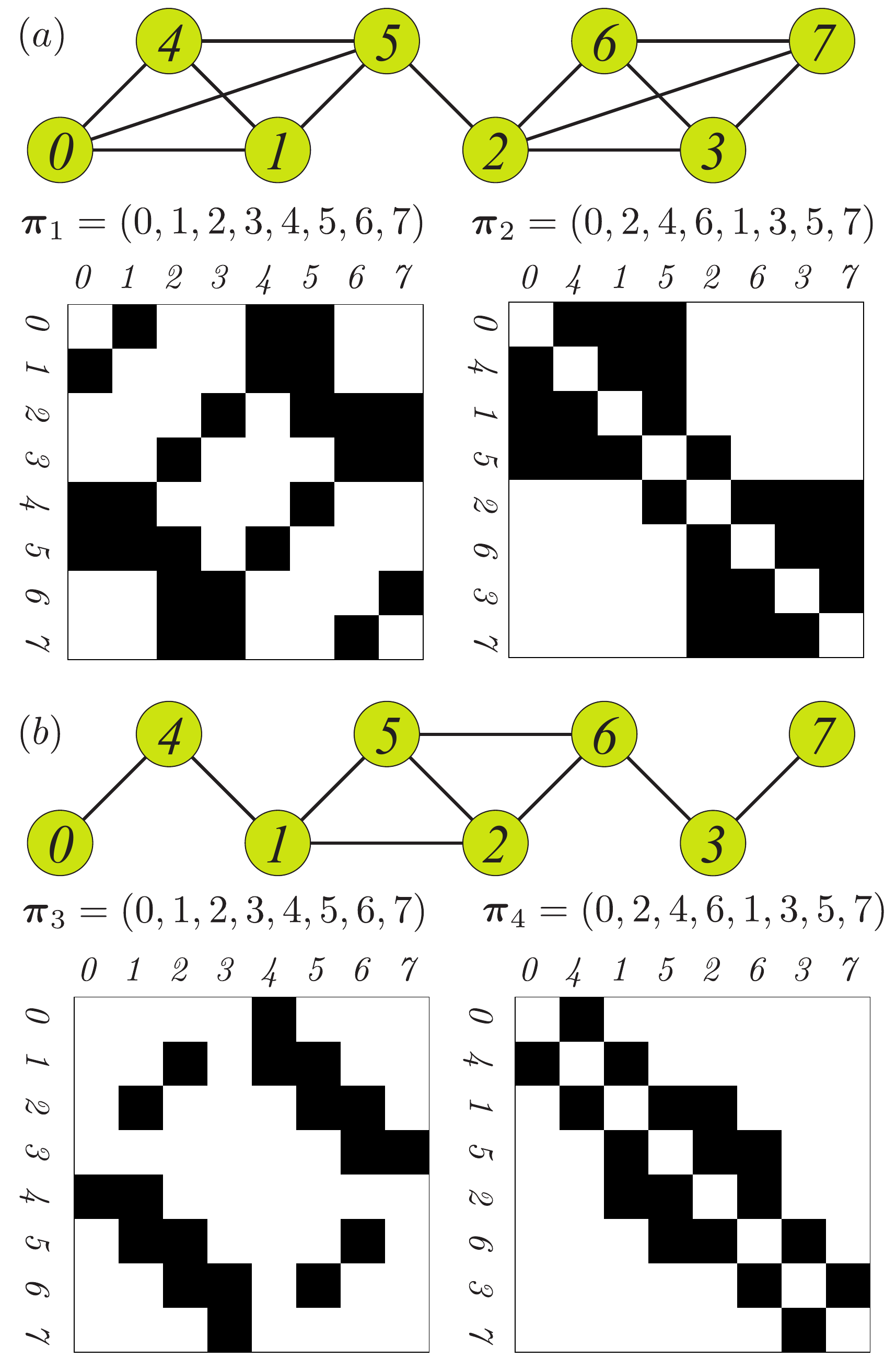}
  \caption{
  Networks with (a) a community structure and (b) a banded structure. 
 Each network is visualized using adjacency matrices with different alignment.
 We define $\ket{\pi}$ in Sec.~\ref{sec:Definition}.
 For both networks, we align matrix elements according to two vertex sequences ($\ket{\pi}_1$ and $\ket{\pi}_2$ for (a), and $\ket{\pi}_3$ and $\ket{\pi}_4$ for (b)), where vertex $i$ corresponds to the $\pi_i$th column (or row) of the adjacency matrix.
  The $(\pi_i,\pi_j)$ element is filled with black when vertices $i$ and $j$ are connected.
  The numbers attached to the matrices represent the raw indices of the vertices.}
  \label{fig:SLS}
\end{figure}

The alignment of adjacency matrix elements, or the ordering of vertices, is critical for the visual identification of macroscopic structures in networks \cite{jeub2015think, sales2007extracting}.
Let us demonstrate two examples.
In Fig.~\ref{fig:SLS}(a), the network has a community structure.
Whereas the community structure is not visible when vertices are not carefully aligned as shown in the left matrix, the adjacency matrix exhibits a nearly block-diagonal structure when vertices are properly aligned as shown in the right matrix.
In Fig.~\ref{fig:SLS}(b), the network has a banded structure.
Again, the banded structure is not visible in the left matrix.
However, when vertices are properly aligned as shown in the right matrix, nonzero elements are rendered to concentrated around the diagonal part.

Importantly, a network may exhibit both community and banded structures simultaneously. 
However, as demonstrated in Sec.~\ref{sec:Results}, previous linear arrangement methods often fail to capture community structures, whereas they allow for the identification of banded structures. 
In this study, we propose a linear arrangement method that allows us to capture community structures in addition to banded structures. 
It is an inference method based on the {\MLE} in which we use the ordered random graph model (ORGM) \cite{Kawamoto2021} as a network generative model.

The rest of the paper is organized as follows: In Sec.~\ref{sec:Definition}, we introduce the basic terminology and related work.
Following the introduction of the {\ORGM} in Sec.~\ref{sec:ORGM}, we formulate the inference method and a greedy algorithm for the optimization of vertex sequence in Sec.~\ref{sec:MLE}. 
Section~\ref{sec:Results} describes the results of the application of the proposed algorithm to synthetic networks and real-world datasets.
Finally, Sec.~\ref{sec:Discussion} is devoted to a discussion of the proposed inference method.

\section{Definitions and related work \label{sec:Definition}}
Let $G = (V, E)$ be a network, where $V$ ($|V| = N$) is the set of vertices and $E$ ($|E| = M$) is the set of edges. 
Throughout this study, we consider undirected networks without multi-edges and self-loops. 
We denote the ordered set of raw vertex indices as $\{0, 1,\dots, N-1\} =: \mathcal{I}$. 
We assign each vertex to a raw vertex index $i\in \mathcal{I}$.
We define the vertex sequence $\ket{\pi} = \left( \pi_0, \pi_1, \cdots, \pi_{N-1} \right)$ by a permutation of the raw vertex indices $\mathcal{I}$, where we use one-line notation \cite{sagan2013symmetric,one_line_notation} to represent the permutation.
For example, the raw vertex index $i \in \mathcal{I}$ is mapped to $\pi_{i} \in \mathcal{I}$ by $\ket{\pi}$.
We denote the inferred vertex sequence by $\hat{\ket{\pi}}$.
We let $\mat{A}$ be the adjacency matrix of a network, where $A_{ij}=1$ if there exists an edge between vertices $i$ and $j$, and $0$ otherwise.
Herein, $\mat{A}$ is a symmetric and binary matrix in which the diagonal elements are all zero. 

Matrix reordering is considered in a variety of contexts.
In the {\MLA} problem, we search for the optimal vertex sequence that minimizes the following cost function \cite{Harper1964}:
\begin{align}
\sum_{i,j=0}^{N-1} A_{ij} | \pi_i - \pi_j |. \label{costfunction-1}
\end{align}
The exact minimization of Eq.~(\ref{costfunction-1}) is $\mathcal{NP}$-complete \cite{Garey1976}, and several lower and upper bounds of the cost function and associated vertex ordering have been discussed in the literature \cite{Juvan1992,Petit2003,Rao2005,Devanur2006,Caprara2011,seitz2010contributions}.
Reordering of symmetric matrices is considered in the context of envelope reduction problem for the efficient computation of sparse matrices.
Several methods to reduce the size of the envelope, which is defined as a banded region around the diagonal part involving all nonzero elements in \cite{Barnard95}, have been proposed \cite{Cuthill1969reducing,Juvan1992,Barnard95,George1997,Alan1981Computer}.
Reordering of rectangular matrices has been established as a seriation problem, which has its origin in archaeology \cite{Robinson1951,Liiv2010}.
For example, the so-called consecutive ones problem considers the reordering of binary matrix such that the ones in each row are aligned consecutively \cite{Kendall1969,Atkins98,Fogel_NIPS2013,Vuokko_SIAM2010}.
Ranking of vertices is studied as a reordering problem of directed networks \cite{brin1998anatomy, de2018physical}.
Clustering methods associated with the ranking are reported in \cite{Peixoto2022Ordered,letizia2018resolution,iacovissi2022interplay}.
Recently, the reordering problem is also studied using neural networks \cite{watanabe2021autoll,kwon2022deep}.
The ordering of vertices can also be considered in the context of latent space network model, as a way of one-dimensional embedding \cite{newman2018networks}.

In this work, we employ two of the envelope reduction algorithms as a comparison with our method; spectral ordering and {\RCM}.
Spectral ordering \cite{Juvan1992,Barnard95,George1997,DingHe2004} employs the eigenvector of the Laplacian matrix to find a vertex sequence that approximately minimizes the cost function
\begin{align}
\sum_{i,j=0}^{N-1} A_{ij} \left( \pi_i - \pi_j \right)^2. \label{costfunction-2}
\end{align}
We sort vertices in the ascending (or descending) order of the eigenvector associated with the second-smallest eigenvalue of the Laplacian matrix.
We employ the normalized Laplacian $\mathcal{\mat{L}} = \mat{D}^{-1/2} \left(\mat{D}-\mat{A}\right) \mat{D}^{-1/2}$, where $\mat{D}$ is the degree matrix defined by $\mat{D}=\mathrm{diag}\left( d_0, d_1, \cdots, d_{N-1}\right)$ and $d_i:=\sum_{j} A_{ij}$ is the degree.
The {\RCM} \cite{Cuthill1969reducing,Alan1981Computer,golub2013matrix} heuristically finds a vertex sequence.
This algorithm permutes vertices using the breadth first search.
See \cite{golub2013matrix} for a detailed explanation.

\section{Ordered random graph model and likelihood function \label{sec:ORGM}}

\subsection{Ordered random graph model}
\begin{figure}[t!]
  \centering
  \includegraphics[width=0.99\columnwidth]{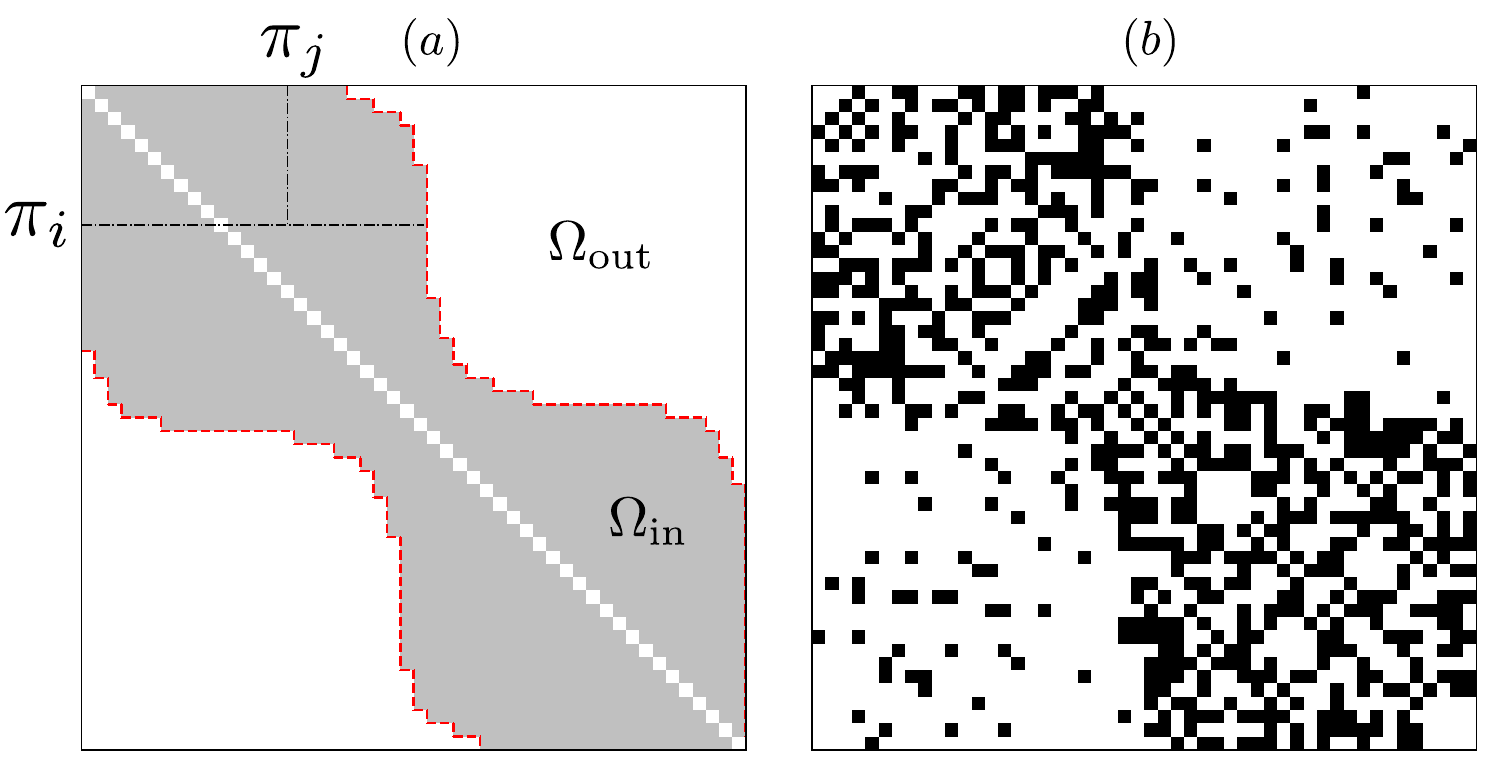}
  \caption{
  	Schematic of the {\ORGM}. 
	(a) The adjacency matrix is divided into the shaded region close to the diagonal ($\Omega_{\mathrm{in}}$) and the rest ($\Omega_{\mathrm{out}}$) by the envelope function $F$ represented by red dashed lines.
	(b) An example of a network generated by (a) with $p_{\mathrm{in}}=0.5$ and $p_{\mathrm{out}}=0.1$.
	}
  \label{fig:Schematic_ORGM}
\end{figure}

The {\ORGM} \cite{Kawamoto2021} is a random graph model that has heterogeneous connection probabilities between vertices based on the planted (or preassigned) vertex sequence $\ket{\pi}$. 
After the vertices are reordered based on $\ket{\pi}$, as illustrated in Fig.~\ref{fig:Schematic_ORGM}(a), we divide the upper-right elements of the adjacency matrix into two regions, $\Omega_{\mathrm{in}}$ and $\Omega_{\mathrm{out}}$. 
These regions are separated by an envelope function $F(i) \in \mathcal{I}$, which is a discrete function with respect to the matrix row index $i \in \mathcal{I}$, i.e., 
\begin{align}
\Omega_{\mathrm{in}} &= \left\{ (\pi_{i}, \pi_{j}) \,|\, |\pi_{i} - \pi_{j}| \le F(\pi_{i}), \, \pi_{i}<\pi_{j} \right\}, \notag\\
\Omega_{\mathrm{out}} &= \left\{ (\pi_{i}, \pi_{j}) \,|\, |\pi_{i} - \pi_{j}| > F(\pi_{i}), \, \pi_{i}<\pi_{j} \right\}. \label{Envelope_discrete}
\end{align}
Note that the envelope function is distinct from the envelope in the envelope reduction problem.

The vertices $i$ and $j$ are connected with probability $p_{\mathrm{in}}$ when the vertices satisfy $(\pi_{i}, \pi_{j})\in \Omega_{\mathrm{in}}$ and are connected with probability $p_{\mathrm{out}}$ when the vertices satisfy $(\pi_{i}, \pi_{j})\in \Omega_{\mathrm{out}}$. 
If we set $p_{\mathrm{in}} > p_{\mathrm{out}}$, vertices with close vertex indices are more likely to be connected.
The envelope function $F$ defines whether the vertices are deemed to be ``close.'' 
Therefore, as illustrated in Fig.~\ref{fig:Schematic_ORGM}(b), the {\ORGM} can generate networks with a complex structure that is not restricted to a banded structure, which is referred to as the sequentially local structure \cite{Kawamoto2021}.

The probability distribution for the set of the adjacency matrix elements $\{A_{ij}\}_{i<j}$ being generated by the {\ORGM} is expressed as follows:
\begin{align}
P \left( \{A_{ij}\}_{i<j} | \ket{\pi}, p_{\mathrm{in}}, p_{\mathrm{out}}, F \right)
&= \prod_{i < j} J_{ij}^{A_{ij}} \left( 1-J_{ij}\right)^{1-A_{ij}}, \label{ORGM_Bernoulli}
\end{align}
where 
\begin{align}
J_{ij} &= \mathrm{Prob}\left[A_{ij} = 1\right] \notag \\
&=
\begin{cases}
p_{\mathrm{in}} & \mathrm{if}\ (\pi_{i}, \pi_{j}) \in \Omega_{\mathrm{in}}\ \mathrm{or}\ (\pi_{j}, \pi_{i}) \in \Omega_{\mathrm{in}}, \\
p_{\mathrm{out}} & \mathrm{if}\ (\pi_{i}, \pi_{j}) \in \Omega_{\mathrm{out}}\ \mathrm{or}\ (\pi_{j}, \pi_{i}) \in \Omega_{\mathrm{out}}.
\end{cases}
\end{align}
While the probability distribution by the Bernoulli distribution is straightforward, we hereafter assume a Poisson distribution instead as
\begin{align}
P \left( \{A_{ij}\}_{i<j} | \ket{\pi}, p_{\mathrm{in}}, p_{\mathrm{out}}, F \right)
&= \prod_{i < j} \frac{ J_{ij}^{A_{ij}} }{A_{ij}!} \mathrm{e}^{-J_{ij}} \label{ORGM1}
\end{align}
because it is easier to work with especially in the {\MLE}.
Although the Poisson distribution possibly generates multigraphs, in the sparse limit when $J_{ij} = \mathcal{O}(1/N)$ with $N\gg1$, we can assume that $p_{\mathrm{in}}$ and $p_{\mathrm{out}}$ are sufficiently small, and thus Eqs.~(\ref{ORGM_Bernoulli}) and (\ref{ORGM1}) are asymptotically identical.

In the inference problem, the vertex sequence $\ket{\pi}$ is the latent variable to be inferred and we estimate the envelope function $F$ and the connection probabilities, $p_{\mathrm{in}}$ and $p_{\mathrm{out}}$. 
For the maximum-likelihood inference of the vertex sequence $\hat{\ket{\pi}}$, we consider the log-likelihood function corresponding to Eq.~(\ref{ORGM1}): 
\begin{align}
&L(\ket{\pi}, p_{\mathrm{in}}, p_{\mathrm{out}}, F|  \mat{A}) \notag\\
&=\mathop{\sum_{i<j}}_{(\pi_{i}, \pi_{j}) \in \Omega_{\mathrm{in}}\ \mathrm{or}\ (\pi_{j}, \pi_{i}) \in \Omega_{\mathrm{in}}} \left(A_{ij} \log p_{\mathrm{in}} -p_{\mathrm{in}} \right) \notag \\
&+ \mathop{\sum_{i<j}}_{(\pi_{i}, \pi_{j}) \in \Omega_{\mathrm{out}}\ \mathrm{or}\ (\pi_{j}, \pi_{i}) \in \Omega_{\mathrm{out}}} \left(A_{ij} \log p_{\mathrm{out}} -p_{\mathrm{out}}\right) \label{ORGM2},
\end{align}
where we neglected the terms that are independent of the vertex sequence $\ket{\pi}$ and model parameters $p_{\mathrm{in}}$, $p_{\mathrm{out}}$, and $F$.
The {\MLE} of the {\ORGM} for a given adjacency matrix $\mat{A}$ is defined as 
\begin{align}
\mathop{\mathrm{argmax}}_{\ket{\pi}, p_{\mathrm{in}}, p_{\mathrm{out}}, F} L(\ket{\pi}, p_{\mathrm{in}}, p_{\mathrm{out}}, F|  \mat{A}) . \label{MLE_ORGM1}
\end{align}
The exact solution of Eq.~(\ref{MLE_ORGM1}) would be obtained by considering all possible vertex sequences, the discrete integer values of envelope function for each row, and connection probabilities. 
However, this is extremely expensive computationally. 

\begin{figure}[t!]
  \centering
  \includegraphics[width=0.8\columnwidth]{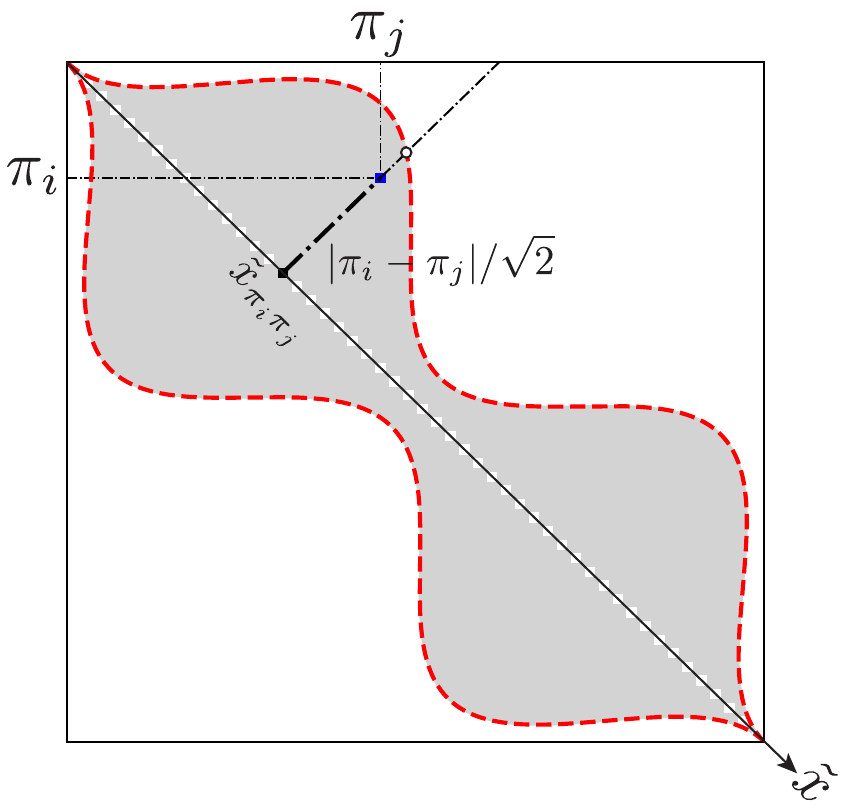}
  \caption{
  	Schematic of the {\ORGM} parametrized by the continuous envelope function $\tilde{b}(\tilde{x})$ ($b(x)$) represented by a red dashed line, which is the boundary of the shaded $\Omega_{\mathrm{in}}$ and unshaded regions $\Omega_{\mathrm{out}}$. 
	}
  \label{fig:Schematic_Jmatrix}
\end{figure}

To efficiently solve the {\MLE}, we modify the definition of the envelope function $F$. 
Instead of a discrete envelope function on the matrix, as illustrate in Fig.~\ref{fig:Schematic_Jmatrix}, we treat the matrix as a two-dimensional plane such that $(i,j)$ element of the matrix corresponds to the $(i,j)$ coordinate and consider a continuous envelope function $\tilde{b}(\tilde{x})$ on this plane. 
We let the upper-left corner of the plane be the origin, $(0,0)$. 
Given a point on the diagonal line, $\tilde{x}$ ($0\le \tilde{x} \le \sqrt{2}\left(N-1\right)$) is the distance between $(0,0)$ and the point, while $\tilde{b}(\tilde{x})$ represents the distance from the point to the boundary of $\Omega_{\mathrm{in}}$ and $\Omega_{\mathrm{out}}$ in the direction perpendicular to the diagonal line. 
When coordinate $(\pi_i,\pi_j)$ is projected to the diagonal line, the distance between the origin and projected point is $\tilde{x}_{\pi_i \pi_j}$, where $\tilde{x}_{ij}=(i+j)/\sqrt{2}$. 
The distance between $(\pi_i,\pi_j)$ and the projected point is $|\pi_i-\pi_j|/\sqrt{2}$. 
Using the continuous envelope function, we redefine the regions $\Omega_{\mathrm{in}}$ and $\Omega_{\mathrm{out}}$ as
\begin{align}
\Omega_{\mathrm{in}} &= \left\{ (\pi_{i}, \pi_{j}) \,|\, |\pi_i-\pi_j| \le \sqrt{2}\tilde{b}(\tilde{x}_{\pi_i \pi_j}), \, \pi_{i}<\pi_{j} \right\}, \notag\\
\Omega_{\mathrm{out}} &= \left\{ (\pi_{i}, \pi_{j}) \,|\, |\pi_i-\pi_j| > \sqrt{2}\tilde{b}(\tilde{x}_{\pi_i \pi_j}), \, \pi_{i}<\pi_{j} \right\}. \label{Envelope_0}
\end{align}

Specifically, we define the continuous envelope function $\tilde{b}(\tilde{x})$ as a superposition of sine-squared functions:
\begin{align}
\tilde{b}(\tilde{x}) = \sum_{k=1}^{K} a_{k} \sin^{2}\left( \frac{\pi k}{  \sqrt{2} \left(N-1\right) } \tilde{x} \right), 
\end{align}
where $K$ is a hyperparameter specifying the limit of complexity of the envelope function, which is treated as a fixed constant.

Hereafter, we rescale $\tilde{b}(\tilde{x})$, $\tilde{x}$, and $\tilde{x}_{ij}$ such that
\begin{align}
b(x)=\sqrt{2}\tilde{b}(\tilde{x}),\quad x=\frac{\tilde{x}}{\sqrt{2}},\quad x_{ij}=\frac{\tilde{x}_{ij}}{\sqrt{2}}=\frac{i+j}{2}.
\end{align}Therefore, 
\begin{align}
\Omega_{\mathrm{in}} &= \left\{ (\pi_{i}, \pi_{j}) \,|\, |\pi_i-\pi_j| \le b(x_{\pi_i \pi_j}), \, \pi_{i}<\pi_{j} \right\}, \notag\\
\Omega_{\mathrm{out}} &= \left\{ (\pi_{i}, \pi_{j}) \,|\, |\pi_i-\pi_j| > b(x_{\pi_i \pi_j}), \, \pi_{i}<\pi_{j} \right\}, \label{Envelope}
\end{align}
\begin{align}
b(x) = \sqrt{2} \sum_{k=1}^{K} a_{k} \sin^{2}\left( \frac{\pi k}{ N-1 } x \right),
\end{align}
where $0 \le x \le N-1$.
The coordinate $(x_{ij}, x_{ij})$ represents the projected point of $(i,j)$ on the diagonal line.

The envelope function satisfies that $b(x)=0$ at the both ends of the diagonal line, $x=0$ and $x = N-1$.
The envelope function is parametrized by the set of real-valued coefficients $\{ a_{k} \} := \{a_{1}, \dots, a_{K} \}$.
We restrict the set $\{a_k\}$ such that the envelope function should not leak out of the upper-triangle of the matrix; that is $0 \leq b(x)\leq \min\{2x, 2(N-1-x)\}$ for $0\leq x\leq N-1$.

This modeling of the envelope function enables us to capture the community structures from the inferred vertex sequence.
Note that, while the number of superpositions $K$ may represent the number of groups, as demonstrated in Sec.~\ref{sec:Results}, it is not always the case. 
Although the inference will be computationally infeasible when $K$ is considerably large, empirically, we found that $K \le 2$ is often sufficient to identify community structures. 

By using the continuous envelope function and the Heaviside step function
\begin{align}
\Theta(x)
&=
\begin{cases}
1 & \mathrm{if}\ x>0, \\
0 & \mathrm{otherwise},
\end{cases}
\end{align}
the log-likelihood function Eq.~(\ref{ORGM2}) is expressed as
\begin{align}
&L(\ket{\pi}, p_{\mathrm{in}}, p_{\mathrm{out}},  \{ a_{k} \}|  \mat{A}) \notag\\
&= 
 \left( \log p_{\mathrm{in}} - \log p_{\mathrm{out}} \right) \, \sum_{i<j} A_{ij} \Theta\left( b(x_{\pi_i\pi_j}) - |\pi_i-\pi_j| \right) \notag\\
&\hspace{10pt} - \left( p_{\mathrm{in}} - p_{\mathrm{out}} \right) \, \sum_{i<j} \Theta\left( b(x_{ij}) - |i-j| \right) \notag\\
&\hspace{10pt} + M \, \log p_{\mathrm{out}} - \frac{N(N-1)}{2} \, p_{\mathrm{out}}. \label{ORGM3}
\end{align}
In the second term of Eq.~(\ref{ORGM3}), we used the fact that $\pi_{i}$ and $\pi_{j}$ can be replaced with $i$ and $j$ because this term is invariant under the permutation of vertices.
As stated in Sec.~\ref{sec:SequenceInference}, this formulation renders the update of the vertex sequence more efficient.

\RestyleAlgo{ruled}
\SetKwComment{Comment}{$\triangleright$\ }{}
\SetKwInOut{Parameter}{Parameters}
\begin{algorithm*}
\caption{The pseudocode of the {\MLE} of the vertex sequence $\hat{\ket{\pi}}$}
\label{GreedyPsuedoCode}
\KwIn{$G$}
\KwOut{$\hat{\ket{\pi}}, \hat{p}_{\textrm{in}}, \hat{p}_{\textrm{out}}, \{\hat{a}_k\}$}
\Parameter{$K$, $\beta$, $\eta$, $\epsilon_1$, $\epsilon_2$, $N_s$}
$\hat{\ket{\pi}} \gets \ket{\pi}_{spectral}$ \Comment*[r]{Initialize $\hat{\ket{\pi}}$ by spectral ordering}
$\{\hat{a}_k\} \gets \{a_k\}_{init}$ \Comment*[r]{Initialize $\{\hat{a}_k\}$ with random value}
$ \delta L \gets \infty$ \Comment*[r]{Initialize}
$ L_0 \gets -\infty$ \Comment*[r]{Initialize}
\While {$|\delta L|> \epsilon_1$} {
	$\{ \hat{p}_{\textrm{in}}, \hat{p}_{\textrm{out}} \} \gets \{ p_{\textrm{in}}, p_{\textrm{out}} \}$ \Comment*[r]{Update $\hat{p}_{\textrm{in}}$ and $\hat{p}_{\textrm{out}}$ by Eq.~(\ref{UpdateP}),(\ref{UpdateQ})}
	\While {$|\nabla_{a_k} L_{\beta}|  > \epsilon_2$} {
	\For {$k=1$ to $K$} {
		$\hat{a}_k \gets \hat{a}_k + \eta \frac{\partial L_{\beta}}{\partial a_k}$ \Comment*[r]{Iteratively update $\{\hat{a}_k\}$ by Eq.~(\ref{UpdataA})}
	}
	}
	\For {$s=1$ to $N_s$} {
		Randomly choose a pair of vertices $\{\ell, m\}$ \;
		$\Delta L_{\ell m} \gets \Delta L_{\ell m}$  \Comment*[r]{Calculate $\Delta L$ by Eq.~(\ref{LossVariation})}
		\If{$\Delta L_{\ell m} > 0$} {
			$\{\hat{\pi}_{\ell},\hat{\pi}_m\} \gets \{\hat{\pi}_m,\hat{\pi}_{\ell}\}$ \Comment*[r]{Update $\hat{\ket{\pi}}$ by greedy algorithm}
		}
	}
	$ L \gets L$ \Comment*[r]{Calculate $L$ by Eq.~(\ref{ORGM3})}
	$ \delta L \gets L-L_0$ \;
	$ L_0 \gets L$ \;
}
\end{algorithm*}

\section{Maximum-likelihood estimate \label{sec:MLE}}

In this section, we estimate model parameters as well as the vertex sequence $\hat{\ket{\pi}}$.
We denote the estimated model parameters as $\hat{p}_{\mathrm{in}}$, $\hat{p}_{\mathrm{out}}$, and $\{ \hat{a}_{k} \}$.
We iteratively conduct the inference of the vertex sequence together with model-parameter learning until convergence.
We show the pseudocode of this iterative process in Algorithm~\ref{GreedyPsuedoCode}.
It is usually infeasible to find the exact vertex sequence that maximizes the log-likelihood function by searching all possible solutions.
Thus, we employ a greedy heuristic to infer the vertex sequence.
For the efficient search of the optimal solution, we use the results of spectral ordering \cite{DingHe2004} as the initial vertex sequence, where nonzero elements have already been aligned near diagonal elements. 
We continue iteration until the variation of $L$ gets smaller than a threshold $\epsilon_1$.
To avoid being trapped into a local optimum, we conduct the entire optimization process several times with randomly chosen starting points with respect to  $\{a_k\}$.

\subsection{Parameter learning}\label{sec:ParameterLearning}
Given an inferred vertex sequence $\hat{\ket{\pi}}$, we update $\hat{p}_{\mathrm{in}}$ and $\hat{p}_{\mathrm{out}}$ based on the saddle-point conditions of $L$. As a result, we obtain the following updating equations:
\begin{align} 
    \hat{p}_{\mathrm{in}} &= \label{UpdateP} \frac{
    \sum_{i<j} A_{ij} \Theta\left( b(x_{\hat{\pi}_i \hat{\pi}_j}) - |\hat{\pi}_i-\hat{\pi}_j| \right) 
    }{
    \sum_{i<j} \Theta\left( b(x_{ij}) - |i-j| \right)
    },\\
    \hat{p}_{\mathrm{out}} &= \label{UpdateQ} \frac{
    \sum_{i<j} A_{ij} \Theta\left( -b(x_{\hat{\pi}_i\hat{\pi}_j}) + |\hat{\pi}_i-\hat{\pi}_j| \right) 
    }{
    \sum_{i<j} \Theta\left( -b(x_{ij}) + |i-j| \right)
    }.
\end{align}

Before considering the update of the set of coefficients $\{ \hat{a}_{k} \}$, we note that the log-likelihood function $L$ is not differentiable with respect to $\{a_k\}$ because the Heaviside step function is discontinuous on the boundary of $\Omega_{\mathrm{in}}$ and $\Omega_{\mathrm{out}}$.
Instead, we approximate the log-likelihood function as follows:
\begin{align}
&L_{\beta}(\ket{\pi}, p_{\mathrm{in}}, p_{\mathrm{out}},  \{ a_{k} \} |  \mat{A}) \notag\\
&= 
 \left( \log p_{\mathrm{in}} - \log p_{\mathrm{out}} \right) \, \sum_{i<j} A_{ij} \Theta_{\beta}\left( b(x_{\pi_i \pi_j}) - |\pi_i-\pi_j| \right) \notag\\
&\hspace{10pt} - \left( p_{\mathrm{in}} - p_{\mathrm{out}} \right) \, \sum_{i<j} \Theta_{\beta}\left( b(x_{ij}) - |i-j| \right) \notag\\
&\hspace{10pt} + M \, \log p_{\mathrm{out}} - \frac{N(N-1)}{2} \, p_{\mathrm{out}}, \label{ORGM4}
\end{align}
where $\beta$ is a hyperparameter and $\Theta_{\beta}(x)$ is a sigmoid function defined as
\begin{align}
\Theta_{\beta}(x)
&=\frac{1}{1+e^{-\beta x}},
\end{align}
which approaches the Heaviside step function as $\beta \to \infty$.

To obtain the saddle-point condition for the set of coefficients $\{ \hat{a}_{k} \}$, we solve the set of equations
\begin{align}
\frac{\partial L_{\beta}}{\partial a_1}=\ \frac{\partial L_{\beta}}{\partial a_2}=\ \cdots=\ \frac{\partial L_{\beta}}{\partial a_K}=0.
\label{eq:saddlepoint_a}
\end{align}
The derivative $\partial L_{\beta} / \partial a_k$ is expressed as
\begin{widetext}
\begin{align}
\frac{\partial L_{\beta}}{\partial a_k}
=   
-\frac{\log p_{\mathrm{in}} - \log p_{\mathrm{out}} }{4} \sum_{i<j} A_{ij} \, 
\frac{\sqrt{2} \beta \sin^2 \left( \frac{\pi k}{N-1} x_{\pi_i \pi_j} \right)}
    {\cosh^2 \left(\frac{\beta}{2} (b(x_{\pi_i \pi_j}) - |\pi_i-\pi_j|)\right)} 
+\frac{p_{\mathrm{in}} - p_{\mathrm{out}}}{4} \sum_{i<j} \, 
\frac{\sqrt{2} \beta \sin^2 \left( \frac{\pi k}{N-1} x_{ij} \right)}
    {\cosh^2 \left(\frac{\beta}{2} (b(x_{ij}) - |i-j|)\right)}. \label{gradL_ak}
\end{align}
\end{widetext}
Unfortunately, the set of equations Eq.~(\ref{eq:saddlepoint_a}) is implicit with respect to $\{a_k\}$ and, thus, is not easy to solve.
Therefore, we employ the gradient descent method.
That is, we maximize $L_{\beta}$ by iteratively updating each $\hat{a}_{k}$ as
\begin{equation} \label{UpdataA}
\hat{a}_k \leftarrow \hat{a}_k +\eta \frac{\partial L_{\beta}}{\partial a_k},
\end{equation}
where $\eta$ is the learning rate that determines the rate of update.
We iterate the update until the convergence specified by $|\nabla_{a_k} L_{\beta}|=\sqrt{\sum_k (\partial L_{\beta}/\partial a_k)^2}<\epsilon_2$, where $\epsilon_2$ is a predetermined threshold.
For the efficient search of the saddle point, we decrease the learning rate as iteration goes on as $\eta=\eta_0/t$, where $\eta_0$ is a hyperparameter and $t$ is the iteration step.

\begin{figure}[t!]
 \centering
  \includegraphics[width=0.7\linewidth]{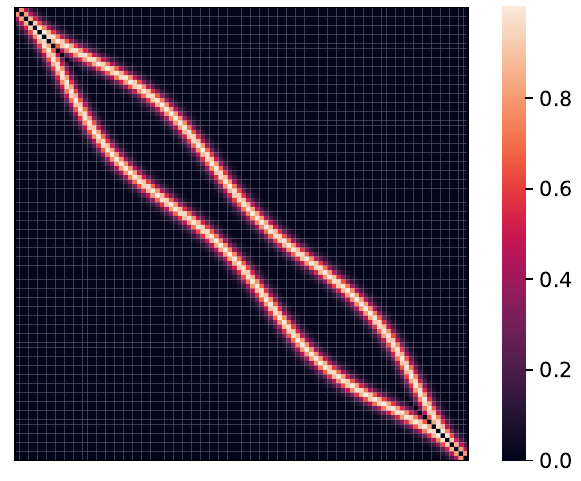}
    \caption{
    Density plot of $\beta/\cosh^2 \left(\beta \left(b\left(x_{ij}\right) - |i-j|\right)/2\right)$ with respect to $i$ and $j$ when $\beta=1$. 
    Here, we set $N=100$, $K=2$, and $a_1=a_2=5$.
    The values are almost equal to zero except for the band around the envelope function.}
    \label{Fig:sparse}
\end{figure}

Even when we use the gradient descent method, the update for $\{ \hat{a}_{k} \}$ is still computationally expensive because we need to evaluate with respect to $N(N-1)/2$ pairs of vertices in the second summation in Eq.~(\ref{gradL_ak}),
\begin{align}
\sum_{i<j} \, 
\frac{\sqrt{2} \beta \sin^2 \left( \frac{\pi k}{N-1} x_{ij} \right)}
    {\cosh^2 \left(\frac{\beta}{2} (b(x_{ij}) - |i-j|)\right)}.
\end{align}
However, when  $\beta$ is sufficiently large, the contribution from a term in this summation is negligible when $|b(x_{ij}) - |i-j|| \gg 0$ because $\beta/\cosh^{2}(\beta x/2) \approx 0$ if $|\beta x| \gg 1$, as shown in Fig.~\ref{Fig:sparse}.
In other words, this term is well-approximated by the sum over the elements near the envelope function:
\begin{align}
&\sum_{d=0}^{2(N-1)} 
\mathop{\sum_{i<j}}_{i+j=d,\ |b(d/2)-|i-j||\le\delta}
\frac{\sqrt{2} \beta \sin^2 \left( \frac{\pi k}{N-1} \frac{d}{2} \right)}{\cosh^2 \left(\frac{\beta}{2} \left(b\left(\frac{d}{2}\right) -  |i-j| \right)\right)},
\end{align}
where $\delta$ is the width of the band around the envelope function in which we count the elements; we set $\delta$ such that $\beta/\cosh^2 \left(\beta \delta /2\right)<10^{-6}$.
The number of pairs of vertices to be taken into account is now reduced to $\mathcal{O}(N)$.

\subsection{Inference of the vertex sequence}\label{sec:SequenceInference}

We solve for the optimal vertex sequence by randomly choosing a vertex pair and exchanging their indices when the log-likelihood function increases.
Considering all possible pairs is computationally expensive when the network size is large because the number of pairs increases with $\mathcal{O}(N^2)$.
We then randomly choose $N_s = n_s N$ pairs of vertices for each iteration, where $n_s$ is a hyperparameter.
When exchanging the indices $\hat{\pi}_{\ell}$ and $\hat{\pi}_m$, the variation of the log-likelihood function is given by
\begin{widetext}
\begin{align}
\Delta L_{\ell m} 
= \notag
\left( \log \hat{p}_{\mathrm{in}} - \log \hat{p}_{\mathrm{out}} \right) 
&\left(
 \mathop{\sum_{j > \ell}}_{j \neq m} \mat{A}_{\ell j} 
 \left( \Theta \left( b \left( x_{\hat{\pi}_{m}\hat{\pi}_{j}} \right) - |\hat{\pi}_{m}-\hat{\pi}_{j}| \right)  - \Theta \left( b \left( x_{\hat{\pi}_{\ell} \hat{\pi}_{j}} \right) - |\hat{\pi}_{\ell} - \hat{\pi}_{j}| \right) \right) 
 \right. \\
 & \left. +  \mathop{\sum_{i > m}}_{i \neq \ell} \mat{A}_{mi} \left( \Theta \left( b \left( x_{\hat{\pi}_i \hat{\pi}_{\ell}} \right) - |\hat{\pi}_i-\hat{\pi}_{\ell} | \right) - \Theta \left( b \left( x_{\hat{\pi}_i \hat{\pi}_m} \right) - |\hat{\pi}_i-\hat{\pi}_m| \right) \right)
 \right)
. \label{LossVariation}
\end{align}
\end{widetext}
Here, we used the fact that only the first summation in Eq.~(\ref{ORGM3}) is relevant to the variation of the log-likelihood function in terms of the vertex sequence as we mentioned before.
For each pair of vertices $(\ell ,m)$, if $\Delta L_{\ell m}>0$, we update the vertex sequence to be $\tilde{\pi}_{\ell}=\hat{\pi}_m$ and $\tilde{\pi}_m=\hat{\pi}_{\ell}$.

We summarize how this iterative process identifies an optimal solution.
In the optimally aligned matrix, the density of nonzero elements in $\Omega_{\mathrm{in}}$ is high, and the densely aligned nonzero elements are tightly surrounded by the envelope function such that zero elements are likely to be excluded from $\Omega_{\mathrm{in}}$.
The combination of updates of the vertex sequence and the model parameters realizes such an alignment.
First, $\hat{p}_{\mathrm{in}}$ and $\hat{p}_{\mathrm{out}}$ are updated to the densities of nonzero elements in $\Omega_{\mathrm{in}}$ and $\Omega_{\mathrm{out}}$, as shown in Eq.~(\ref{UpdateP}) and (\ref{UpdateQ}), respectively.
Second, we consider the update of the shape of the envelope function $\{\hat{a}_k\}$.
Let us assume that one of the nonzero elements moves from $\Omega_{\mathrm{out}}$ to $\Omega_{\mathrm{in}}$ by altering the envelope function.
In this case, Eq.~(\ref{ORGM3}) shows us that $L$ increases by $(\log \hat{p}_{\mathrm{in}} - \log \hat{p}_{\mathrm{out}})-(\hat{p}_{\mathrm{in}}-\hat{p}_{\mathrm{out}})$.
Therefore, when $\Omega_{\mathrm{out}}$ is sufficiently sparse, the nonzero elements are more likely to be included in $\Omega_{\mathrm{in}}$.
In contrast, when a zero element moves from $\Omega_{\mathrm{in}}$ to $\Omega_{\mathrm{out}}$, $L$ increases by $\hat{p}_{\mathrm{in}}-\hat{p}_{\mathrm{out}}$.
Thus, zero elements are more likely to be excluded from $\Omega_{\mathrm{in}}$.
Finally, $L$ is increased by updating $\hat{\ket{\pi}}$ such that more nonzero elements are included in $\Omega_{\mathrm{in}}$ when $\hat{p}_{\mathrm{in}}>\hat{p}_{\mathrm{out}}$, as observed in Eq.~(\ref{ORGM3}) .
Thus, the vertex sequence is updated so that more nonzero elements are included in $\Omega_{\mathrm{in}}$.

\section{Results}\label{sec:Results}

\begin{figure*}[ht!]
 \centering
  \includegraphics[width=0.9\linewidth]{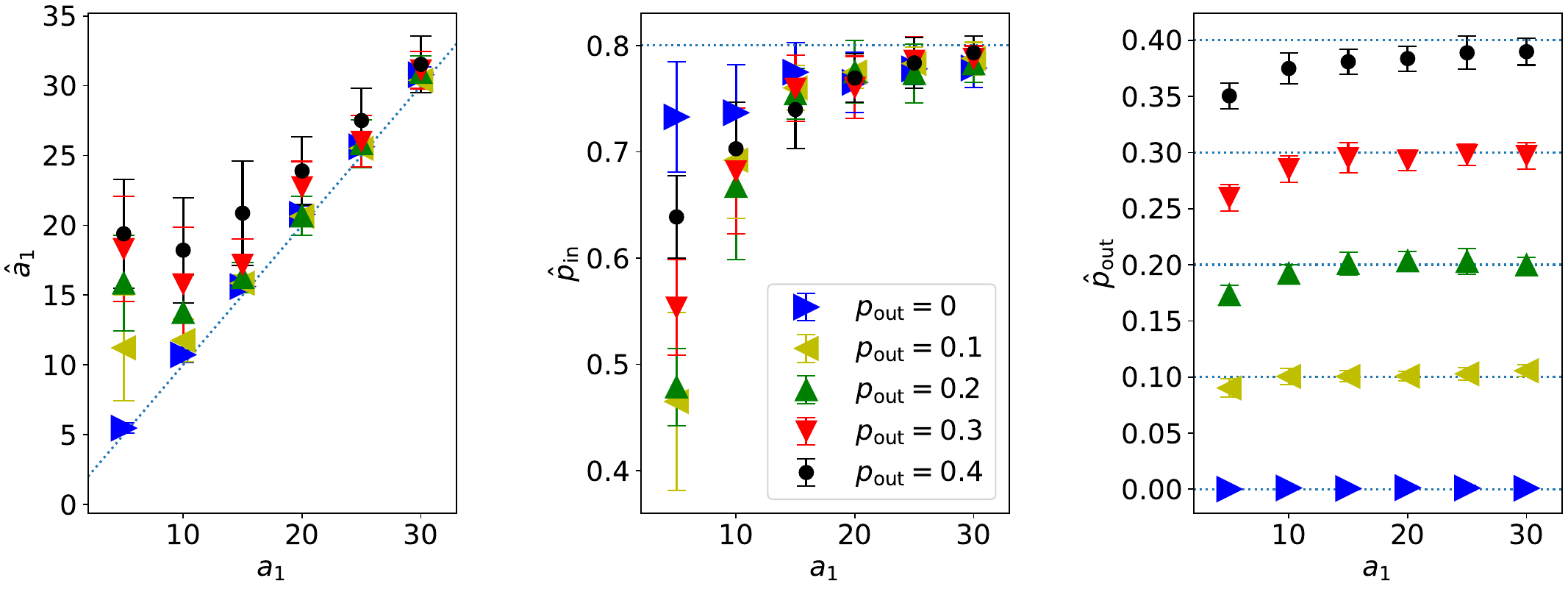}
    \caption{
    The estimated model parameters of the {\ORGM} by the {\MLE} with $K=1$.
    We generate networks by the {\ORGM} with $N=100$, $K=1$, $p_{\mathrm{in}}=0.8$ with various values of $a_1$ and $p_{\mathrm{out}}$.
    The dashed lines represent the cases where the estimated values coincide with the true values of the model.
    Each symbol and bar represent the average and standard deviation over $20$ network instances, respectively. 
    }
    \label{fig:EstimateORGM}
\end{figure*}

\subsection{Synthetic networks}\label{sec:synthetic}

We examine the performance of the {\MLE} of the {\ORGM} by applying to synthetic networks.
In this section, we set the hyperparameters to be $\beta=10$, $\eta_0=0.1$, $\epsilon_1=10^{-6}$, $\epsilon_2=0.1$, $\delta=2$, and $n_s=10$, and we consider $100$ initial states for $\{a_k\}$ and employ the solution with the largest likelihood.

First, we apply the {\MLE} to the networks generated by the {\ORGM} and examine whether the algorithm can correctly infer the model parameters.
We generate networks by setting the model parameters of the {\ORGM} to be $N=100$, $K=1$, $p_{\mathrm{in}}=0.8$ with various values of $a_1$ and $p_{\mathrm{out}}$.
Figure~\ref{fig:EstimateORGM} shows the model parameters estimated by the {\MLE} with $K=1$.
We observe that the estimated model parameters are highly accurate when $p_{\mathrm{out}}=0$.
As $p_{\mathrm{out}}$ increases, the estimated parameters divert from the true values for smaller values of $a_1$.
This experiment implies that the {\MLE} of the {\ORGM} can estimate the model parameters correctly when the number of nonzero elements in $\Omega_{\textrm{in}}$ is large.
In Appendix~\ref{sec:appendix_ORGM}, we investigate how the performance of the estimate gets deteriorated through the experiments with smaller $p_{\mathrm{in}}$.

\begin{figure*}[ht!]
 \centering
  \includegraphics[width=0.99\linewidth]{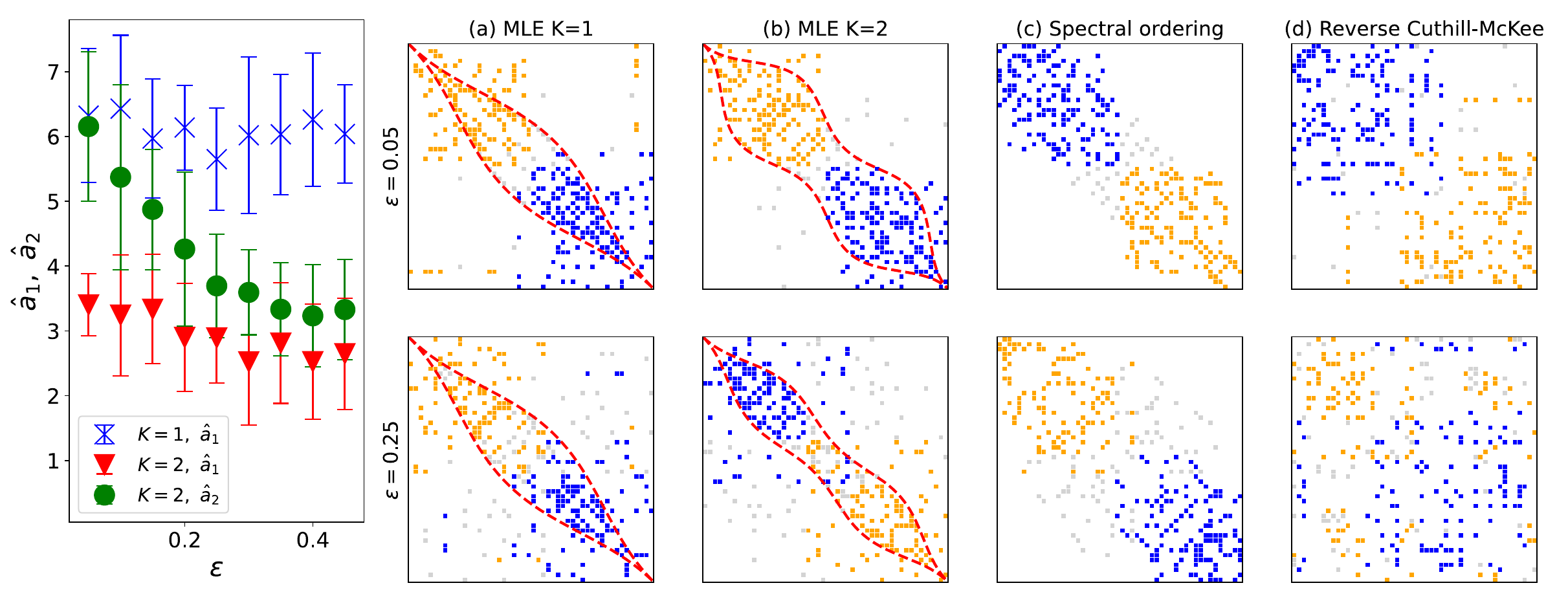}
    \caption{
    (Left) Estimated $\{\hat{a}_k\}$ with the {\MLE} when applied to the networks generated by the {\SBM} with $N=50$, $B=2$ and $c=6$.
    Each symbol and bar represent the average and standard deviation over $20$ network instances, respectively.
    (Right) Heatmaps of adjacency matrices for a single network instance when $\epsilon=0.05$ (upper row) and $\epsilon=0.25$ (lower row).
    The dashed (red) curves represent the envelope function.
    We estimate the vertex sequence with four algorithms: the {\MLE} (MLE) with (a) $K=1$ and (b) $K=2$, (c) spectral ordering, and (d) {\RCM}.
    If vertices $i$ and $j$ are connected and belong to the same group, the matrix element $A_{ij}$ is filled with a dark color (blue) or pale color (yellow) depending on the preassigned group label.
    When $i$ and $j$ are connected and belong to the different groups, the matrix element $A_{ij}$ is filled with an even paler color (gray).
    }
    \label{HeatmapComparison_SBM_B=2}
\end{figure*}

\begin{figure}[t!]
 \centering
  \includegraphics[width=0.99\linewidth]{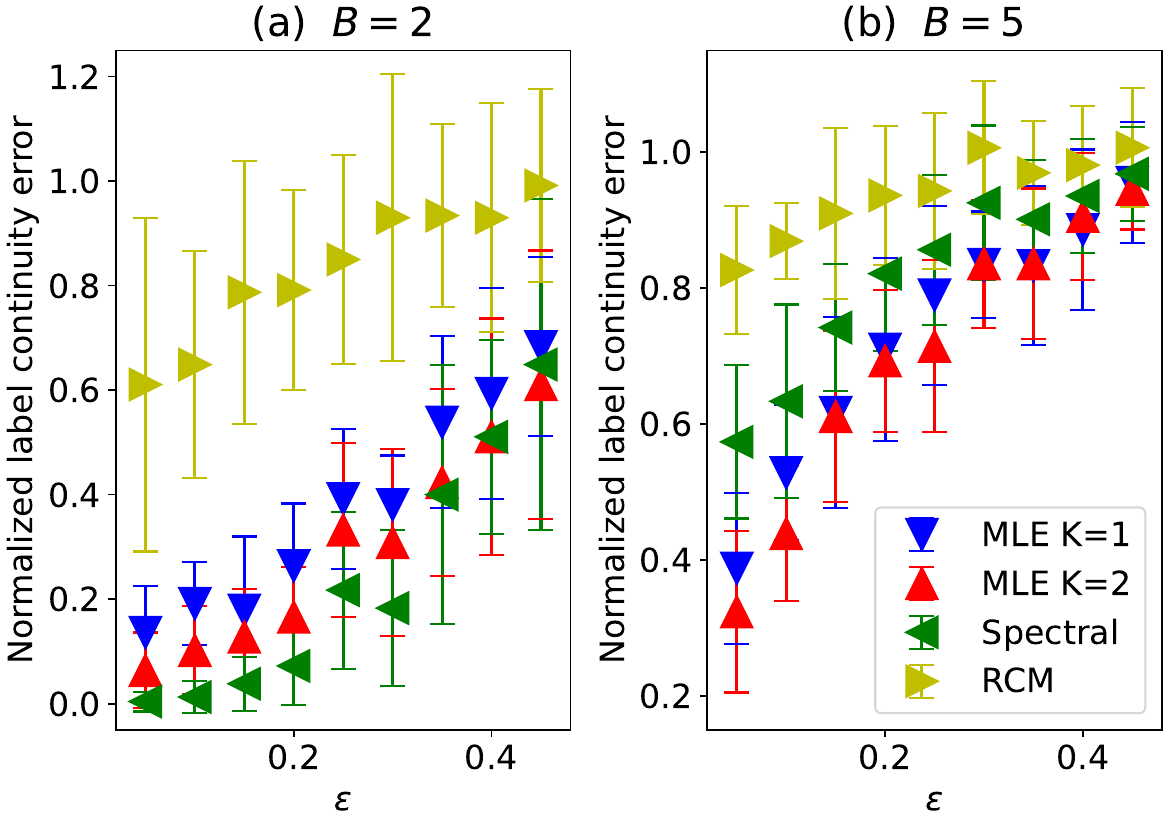}
    \caption{
    The normalized label continuity error between the estimated vertex sequence and the preassigned group labels of the {\SBM} using the {\MLE} (MLE) (K=1), {\MLE} (MLE) (K=2), spectral ordering (Spectral), and {\RCM} (RCM).
    The number of groups are (a) two and (b) five respectively.
    In both cases, we set $N=50$ and $c=6$, and each symbol and bar represent the average and standard deviation over $20$ network instances, respectively.
    }
    \label{EstimateSBM_nlce}
\end{figure}

Next, we examine to what extent our algorithm can infer the community structure. 
To this end, we apply the {\MLE} to the networks with community structure generated by stochastic block model ({\SBM}) \cite{holland1983stochastic}.
The {\SBM} generates a network as follows.
We first define the number of group $B$, and assign the group memberships to the vertices $\ket{\sigma} = \left( \sigma_0, \sigma_1, \cdots, \sigma_{N-1} \right), \sigma_i \in \{0,1,\cdots,B-1\}$.
Then, vertices $i$ and $j$ are connected with the probability $q_{\mathrm{in}}$ when $\sigma_i=\sigma_j$ and with the probability $q_{\mathrm{out}}$ when $\sigma_i\neq \sigma_j$.
Hereafter, we assume that the size of each group is equal.
The strength of community structure is represented by the ratio of the probabilities $\epsilon=q_{\mathrm{out}}/q_{\mathrm{in}}$.
The adjacency matrix exhibits a block-diagonal structure when $\epsilon=0$ while the network is uniformly random when $\epsilon=1$.
Note that $q_{\mathrm{in}}$ and $q_{\mathrm{out}}$ are uniquely determined when $N$, $B$, $\epsilon$ and the average degree $c=2M/N$ are given.

We generate networks by setting the model parameter of the {\SBM} to be $N=50$ and $c=6$ with various values of $\epsilon$ and we consider $20$ network instances for each $\epsilon$.
We perform the {\MLE} with $K=1$ and $K=2$.
We consider the spectral ordering \cite{DingHe2004} and the {\RCM} \cite{Cuthill1969reducing,Alan1981Computer} for comparison.
Figure~\ref{HeatmapComparison_SBM_B=2} shows the estimated model parameters $\{ \hat{a}_k \}$ of the {\ORGM} and the heatmaps of the adjacency matrix based on the estimated vertex sequence.
We observe that $a_2$ decreases as the strength of community structure becomes weaker when $K=2$.
This result indicates that the envelope function has two peaks corresponding to the groups when $\epsilon$ is smaller, while these peaks decay as $\epsilon$ increases and the community structure becomes weaker.
When $\epsilon=0.05$, the community structure is visible through the heatmap using the maximum-likelihood method with $K=1$. 
Furthermore, we can observe that the network exhibits a mixture of the community and banded structures when $K=2$ is used. 
Note that it is natural that the instances of the {\SBM} also exhibits a banded structure because a sparse uniform random graph often exhibits a banded structure when the vertices are properly aligned \cite{Kawamoto2021}.
The banded structure is also visible in the heatmap using the spectral ordering.
The {\RCM} fails to exhibit the banded structure whereas the community structure is visible.
In contrast, when $\epsilon=0.25$, we cannot find the community structure in all the four heatmaps while we can observe the banded structure in the heatmap using the spectral ordering.

Let us numerically evaluate how the estimated vertex sequence $\hat{\ket{\pi}}$ is consistent with the preassigned group membership $\ket{\sigma}$.
We use normalized label continuity error \cite{kawamoto2022consistency}, which is defined by
\begin{align}
L(\hat{\ket{\pi}},\ket{\sigma}) = 
\frac{ \displaystyle N-B-\sum_{i'=0}^{N-2} \delta \left( \sigma\left( \hat{\pi}^{-1} (i')  \right), \sigma\left( \hat{\pi}^{-1} (i'+1)  \right)  \right)}{N-B-(N-1)/B},
\end{align}
where $\delta\left( x, y \right)$ is the Kronecker delta and $\sigma\left(\hat{\pi}^{-1}(i)\right)$ is the group label of the vertex $i'$ whose vertex index is $\hat{\pi}_{i'} = i$.
The normalized label continuity error takes zero when all vertices in each group are aligned maximally consecutively.
 In other words, the result of ordering is consistent with the group labels.
It takes a higher value when the neighboring vertices often belong to different groups.

In Fig.~\ref{EstimateSBM_nlce}(a), we show the normalized label continuity error for the estimated vertex sequences when the number of groups is two ($B=2$); this is the case we considered in Fig.~\ref{HeatmapComparison_SBM_B=2}.
When $\epsilon$ is small, we find that the vertex sequences from the {\MLE} with $K=2$ and the spectral ordering are highly consistent with the preassigned group labels, while the {\RCM} is less consistent (larger normalized label continuity error).
We consider the {\SBM} with five groups ($B=5$) in Fig.~\ref{EstimateSBM_nlce}(b).
In this case, the vertex sequence of the {\MLE} with both $K=1$ and $K=2$ are more consistent with the group labels than that of the spectral ordering contrary to the case with two groups.
These experiments show that the {\MLE} aligns vertices so that the community structure is visible even when the number of planted peaks in the envelope function $K$ is smaller than $B$.

\subsection{Real-world datasets}\label{sec:realworld}

\begin{table}[t!]
\begin{center}
\begin{tabularx}{\columnwidth}{lccl}
 \hline
   \multirow{2}{*}{Dataset} &   \multirow{2}{*}{$N$} &   \multirow{2}{*}{$M$}  &\multirow{2}{*}{Data description}  \\
   & & &  \\
   \hline             
          Political Books \cite{Newman2006politicalbooks}  &  105 &   441  &  \begin{tabular}{l}Co-purchase network of \\ books on US politics.\end{tabular}   \\     
          Les Mis\'{e}rables   \cite{knuth1993stanford} &  77 &   254 &   \begin{tabular}{l}Character co-appearance\\ network of  \emph{Les Mis\'{e}rables}.\end{tabular}   \\    
          Football  \cite{girvan2002community,Evans2010} &  115 &   613 &  \begin{tabular}{l}Network of American\\ football games between\\ Division IA colleges.\end{tabular}  \\                                  
    \hline 
 \end{tabularx}
 \end{center}
 \caption{Description of empirical datasets. All the data can be loaded from graph-tool~\cite{graph-tool}.}
 \label{tab:dataset}
\end{table}

\begin{figure*}[ht!]
 \centering
  \includegraphics[width=0.9\linewidth]{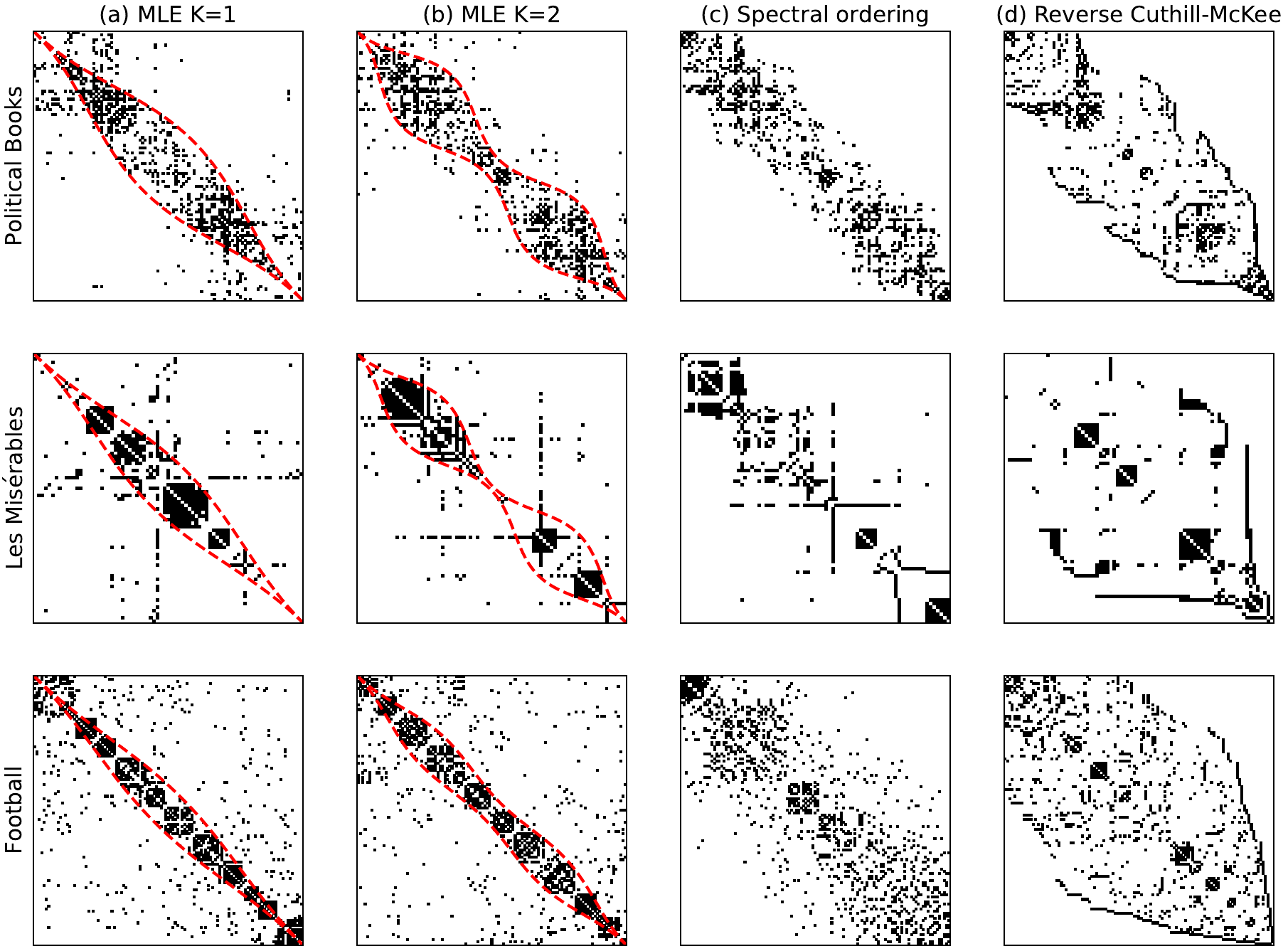}
    \caption{Heatmaps of the adjacency matrices of three real-world datasets ordered by the four algorithms. The matrix element filled with black represents $A_{ij}=1$. The red dashed line represents the estimated envelope function. 
    The top, middle, and bottom panels represent the results for the networks of Political Books, Les Mis\'{e}rables, and Football, respectively. 
    Each column represents the results of different algorithms:  our {\MLE}s of the {\ORGM} with (a) $K=1$ and (b) $K=2$, (c) spectral ordering, and (d) {\RCM}. 
    }
    \label{HeatmapComparison}
\end{figure*}

We demonstrate the performance of the {\MLE} of the {\ORGM} using three real-world network datasets that are often used in network science; the networks of Political Books \cite{Newman2006politicalbooks}, Les Mis\'{e}rables \cite{knuth1993stanford}, and Football \cite{girvan2002community,Evans2010} (Table~\ref{tab:dataset}).

The visualization of the adjacency matrices based on the inferred vertex sequence is shown in Fig.~\ref{HeatmapComparison}. 
We show the results with $K=1$ and $K=2$, and set the parameters to be $\beta=10$, $\eta_0=0.1$, $\epsilon_1=10^{-6}$, $\epsilon_2=0.1$, $\delta=2$, and $n_s=10$. 
For each case, we consider $1000$ initial states for $\{a_k\}$ and employ the solution with the largest likelihood.
Our results are compared with those of the spectral ordering \cite{DingHe2004} and the {\RCM} \cite{Cuthill1969reducing,Alan1981Computer}, which are also shown in Fig.~\ref{HeatmapComparison}.

The {\MLE} allows us to capture the sequentially local structure through the linear arrangement for all three datasets.
The envelope functions, illustrated as red curves, are inferred relatively close to the diagonal elements, and the nonzero elements, illustrated as black elements, are more likely to be placed inside the envelope function for both cases when $K=1$ and $K=2$.
The {\MLE} of the {\ORGM} exhibits a block-diagonal structure more clearly compared with the results of the spectral ordering and the {\RCM}.
We also show the results with $K=3$ and $K=4$ in Appendix~\ref{sec:appendix}, where we observe that larger value of $K$ does not necessarily allow us to capture clear community structures.

The result on the Political Books network with $K=2$ exhibits two groups corresponding to the two peaks of the envelope function. 
We can also confirm that the Political Books network has a sequentially local structure that is a mixture of the community and banded structures. 
For the Les Mis\'{e}rables and Football networks, we can identify several clique-like groups even when $K=2$ is employed. 
In contrast, the spectral ordering and the {\RCM} often focus more on banded structures. 

\begin{figure*}[t]
 \centering
  \includegraphics[width=0.9\linewidth]{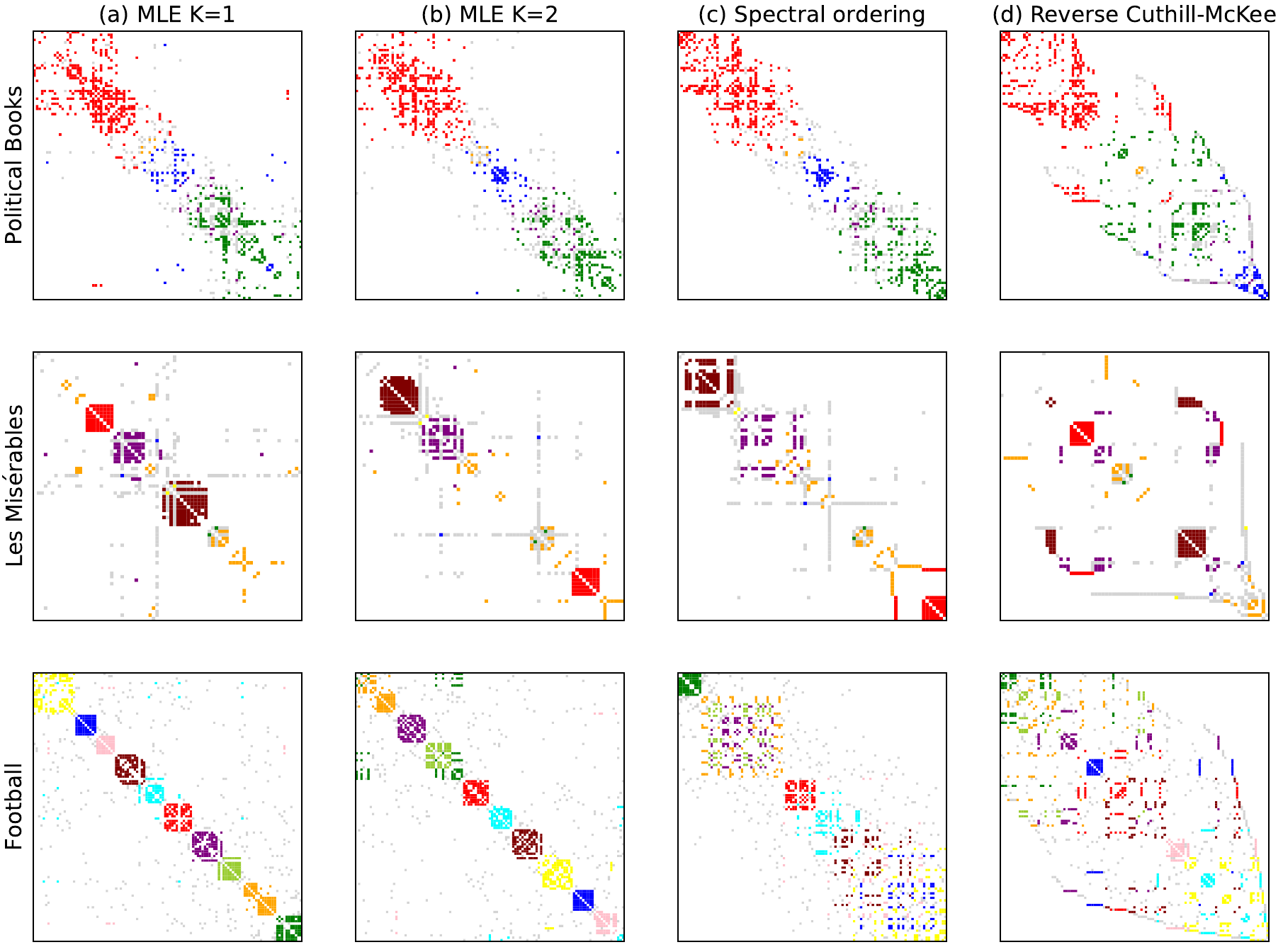}
    \caption{Heatmaps of adjacency matrices that reflect the results of community detection. 
    Matrix elements filled with the same (non-gray) color represent the vertex pairs belonging to the same group, while other nonzero elements are filled with the gray color. 
    For the alignment of matrix elements, we use the same vertex sequence as in Fig.~\ref{HeatmapComparison}.
    }
    \label{HeatmapClustering}
\end{figure*}

We next examine whether the optimized vertex sequences are consistent with an existing community detection method. 
We use the Markov-chain Monte Carlo algorithm based on non-degree-corrected version of the {\SBM} \cite{peixoto2014efficient,peixoto2015model}, which is implemented in \textit{graph-tool} \cite{graph-tool}. 
This algorithm automatically determines the number of groups.
We obtained five groups for the Political Books network, seven for the Les Mis\'{e}rables network, and 10 for the Football network. 
We can visually confirm from Fig.~\ref{HeatmapClustering} that the vertices in the same group are more closely aligned, indicating that the vertex sequence obtained via our linear alignment algorithm is consistent with the results of the community detection algorithm. 

Interestingly, in the Political Books network, the optimal vertex sequence shows that the two large groups, which are respectively represented by red and green, are connected via a small number of vertices, while such vertices in the middle are regarded to be forming another group, which is represented by blue, in the community detection method.
That is, the group in the middle serves as a bridge at the ambiguous boundary of two large groups.
Note that, if we inferred the group labels only and aligned the vertices based on those labels, we would simply conclude that the network has a community structure with two large groups and one small group. 
However, a careful alignment of the vertices allows us to interpret that the small group is in fact a subgraph that smoothly connect the other large groups.
Therefore, we can obtain a better understanding of a dataset by applying both linear alignment and community detection algorithms when the results of these two algorithms are consistent with each other.

\section{Discussion}\label{sec:Discussion}

In this study, we proposed a linear arrangement algorithm based on statistical modeling.
We employed the {\ORGM} as a generative model of networks and performed its {\MLE}. 
Our statistical framework to infer the sequentially local structure provides a new perspective to the community detection in a network.
While we employed the {\MLE} in this work, our framework can be extended to Bayesian inference.

Our modeling is more flexible than the classical linear arrangement methods that are based on a cost function, such as Eq.~(\ref{costfunction-1}), and thus, our method allows us to capture a sequentially local structure that is a mixture of community and banded structures.
For example, in the Football network, the community structure is not observed in the results of the spectral ordering and the {\RCM}.
This is because these methods strongly penalize the nonzero elements far from the diagonal part.
In contrast, our method is not prone to such a problem because it penalizes matrix elements in a more flexible manner.

Our linear arrangement algorithm can detect community structures with ambiguous boundaries, as we observed from the result of the Political Books network (Fig.~\ref{HeatmapClustering}).
In contrast, many of the previous community detection methods perform ``hard clustering'', in which the detected groups have clear boundaries because a group label is assigned to each vertex. 
Such a hard clustering method is not suitable for some networks, and several other flexible methods were proposed in the literature. 
Overlapping community detection methods \cite{ahn2010link, xie2013overlapping} deal with ambiguous boundaries by assigning multiple group labels to each vertex.
Bayesian inference methods \cite{peixoto2014hierarchical,NewmanReinert2016} infer a probability distribution of group assignment for each vertex. 
Although the interpretation of ambiguous boundaries of groups in these methods is not identical, they achieve ``soft clustering'' using a higher-order group membership. 
Our algorithm simply estimates the vertex sequence and allows us to visually identify the fuzziness of community structures without introducing higher-order group memberships.

Another advantage of our approach is that we neither need to give nor infer the number of groups. 
For example, in clustering methods, such as an inference method based on the {\MLE} of the {\SBM} \cite{karrer2011stochastic} and spectral clustering \cite{von2007tutorial}, the number of groups needs to be provided as input. 
In some methods, such as the Bayesian inference, the number of groups is determined in a nonparametric manner \cite{peixoto2014hierarchical,NewmanReinert2016}. 
In contrast, as observed in the results of the Les Mis\'{e}rables and Football networks (Fig.~\ref{HeatmapComparison}), we can visually estimate an appropriate number of groups when the vertices are optimally aligned. 
There are pros and cons to our method. 
Specification of the number of groups is ambiguous compared with the clustering methods. 
On the other hand, our approach is more flexible for detecting the mixture of the community and banded structures because the specification of the number of groups in the clustering methods restricts the structures to be detected.

In our method, it is important that $p_{\mathrm{in}}>p_{\mathrm{out}}$ is satisfied.
When $p_{\mathrm{in}}>p_{\mathrm{out}}$, as described in Sec.~\ref{sec:SequenceInference}, the nonzero elements tend to be included in $\Omega_{\mathrm{in}}$ as the vertex sequence is updated.
In contrast, when $p_{\mathrm{in}}<p_{\mathrm{out}}$, the nonzero elements tend to be excluded from $\Omega_{\mathrm{in}}$ as the vertex sequence is updated.
Conducting the spectral ordering as a preprocess helps us prevent this problem, in addition to the efficient search of the optimal vertex sequence, because the spectral ordering aligns more nonzero elements near the diagonal elements.

We characterized the envelope function as a superposition of the sine-squared functions. 
This modeling makes computation efficient because we have only to tune $\{a_k\}$ and is effective for identifying community structures. 
However, other modelings are also possible, and exploring a better modeling for the envelope function would be a possible research direction.
In addition, identifying the optimal number of sine waves $K$ is another task.
In this study, although we have treated $K$ as a hyperparameter and successfully found the sequentially local structures with $K=1$ and $K=2$, a good model selection method can potentially find an optimal value of $K$.
For example, it can be done by considering a Bayesian framework and adding a prior distribution of $K$.

\begin{figure}[t]
 \centering
  \includegraphics[width=0.9\linewidth]{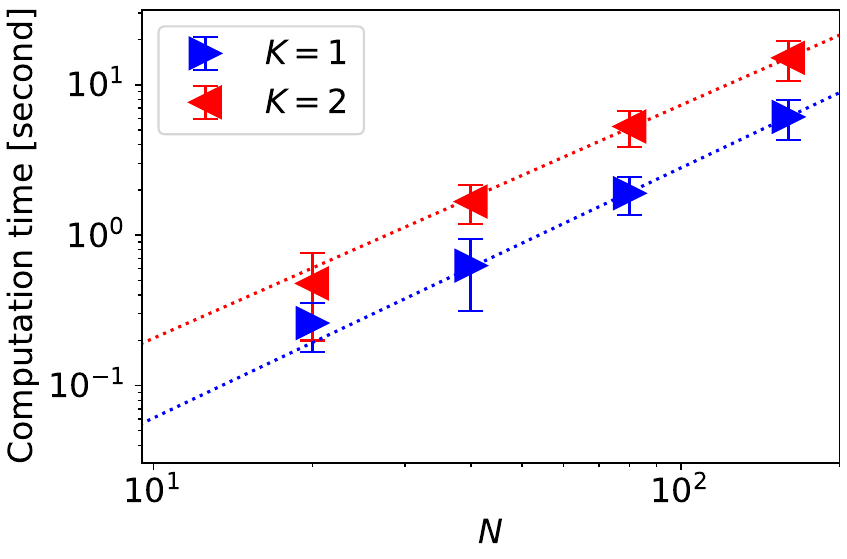}
    \caption{
    Computation time of the {\MLE} for a single initial value of $\{a_k\}$.
    We consider regular random networks with $c=6$, and perform the {\MLE} with $K=1$ and $K=2$.
    Each symbol and bar respectively represent the average and the standard deviation of the $100$ samples of the initial value, and the dotted lines is the fitted curve by $y=ax^b$ using the least squares .
    When $K=1$, $a=1.33\times 10^{-3}$ and $b=1.66$, and when $K=2$, $a=5.85\times 10^{-3}$ and $b=1.55$.
    For the calculation, we use the 2020 Macbook Pro with Apple M1 chip.}
    \label{Fig:MLE_time}
\end{figure}

Finding an optimal vertex sequence generally requires a large computational cost. 
To make the optimization tractable, we restricted the envelope function to a superposition of sine-squared functions in the proposed method.
We also applied the transformation in Eq.~(\ref{ORGM3}) and the approximation in Eq.~(\ref{gradL_ak}) to the log-likelihood function. 
Furthermore, we employed a greedy heuristic to realize a fast optimization.
Despite these treatments, the {\MLE} of the {\ORGM} is still computationally expensive. 
However, our method at least improves the efficiency of the optimization compared with the brute-force method which requires $\mathcal{O}(N!)$ operations.
In Fig.~\ref{Fig:MLE_time}, we show the computation time to find the optimal vertex sequence when we apply the {\MLE} to the regular random networks.
We find that the computation time increases polynomially as the number of vertices increases, with the exponent less than two.
In addition, although the computational cost increases when we use a larger value of $K$, the exponent is almost same between the cases with $K=1$ and $K=2$.

\appendix

\section{\label{sec:appendix_ORGM} Accuracy of the {\ORGM} parameters}

\begin{figure}[t]
 \centering
  \includegraphics[width=0.99\linewidth]{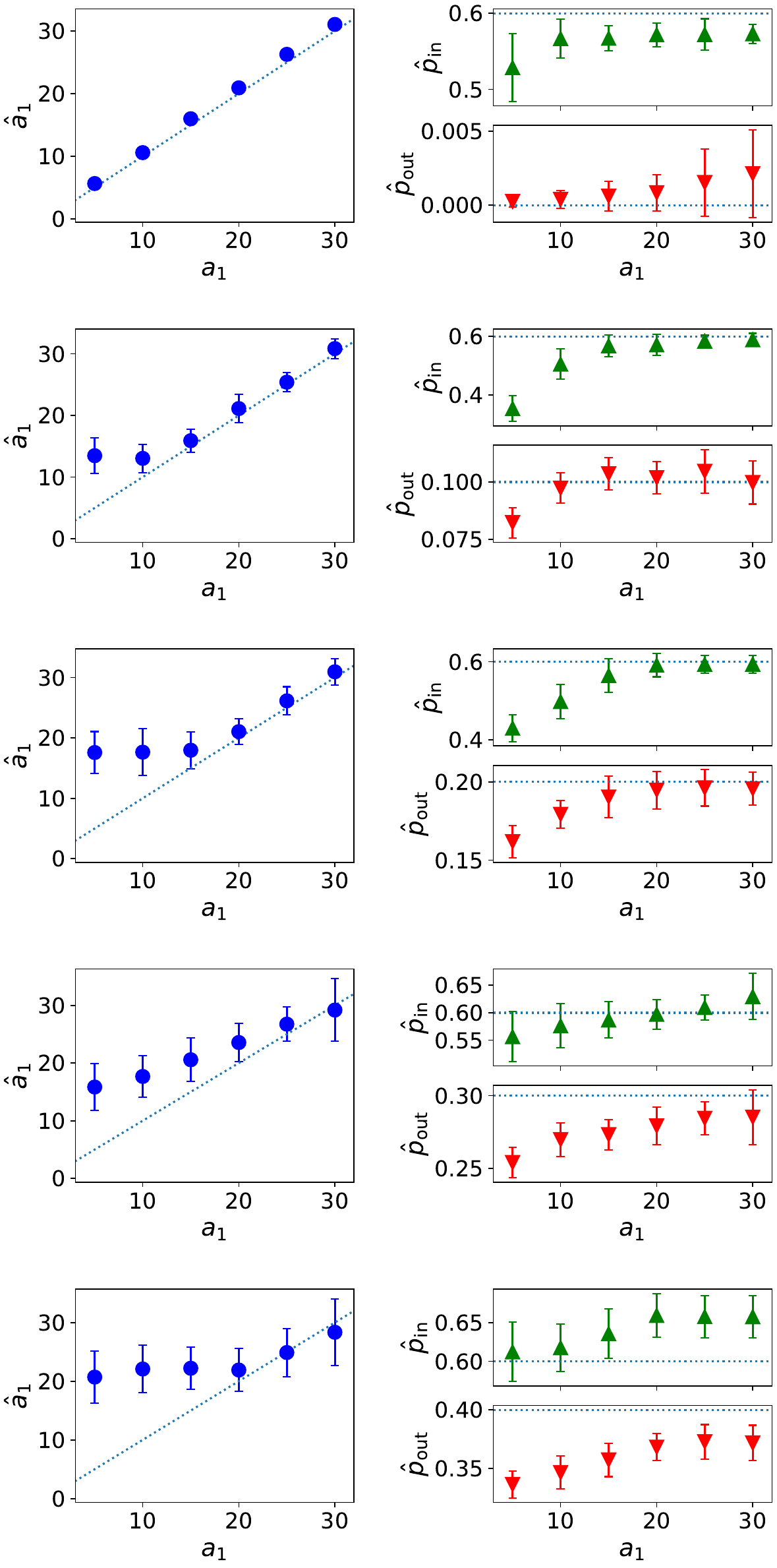}
    \caption{
    The estimated model parameters of the {\ORGM} by the {\MLE} with $K = 1$. 
    The networks are generated by the ORGM with $N = 100$, $K = 1$ and $p_{\mathrm{in}}=0.6$.
    The symbols, bars, and dashed lines are used in the same manner as in Fig.~\ref{fig:EstimateORGM}.}
    \label{EstimateORGM_pin06}
\end{figure}

In the main text, we observed that the {\MLE} correctly infers the true model parameters of the {\ORGM} for larger values of $a_1$ when $p_{\mathrm{in}}=0.8$.
Here, we examine how the performance of the {\MLE} gets deteriorated as we make $p_{\mathrm{in}}$ and $p_{\mathrm{out}}$ closer to those for uniformly random networks.
Figure~\ref{EstimateORGM_pin06} shows the performance of the {\ORGM} parameter estimation.
Herein, we set $p_{\mathrm{in}}=0.6$ for the true parameter and considered various values of $p_{\mathrm{out}}$.
As observed in the case with $p_{\mathrm{in}}=0.8$ in the main text, $a_1$ is overestimated when the true value is small; we confirm that this tendency is enhanced as $p_{\mathrm{in}}$ becomes smaller. 
For the networks with $p_{\mathrm{out}} \leq 0.2$, the estimates $\hat{p}_{\mathrm{in}}$ and $\hat{p}_{\mathrm{out}}$ are accurate whenever $\hat{a}_1$ is accurate. 
However, for the networks with larger values of $p_{\mathrm{out}}$,  $\hat{p}_{\mathrm{in}}$ and $\hat{p}_{\mathrm{out}}$ are not accurate even when $a_1$ and $\hat{a}_1$ apparently coincide, implying that the {\MLE} may have reached its performance limit.

\begin{figure}[b!]
 \centering
  \includegraphics[width=0.99\linewidth]{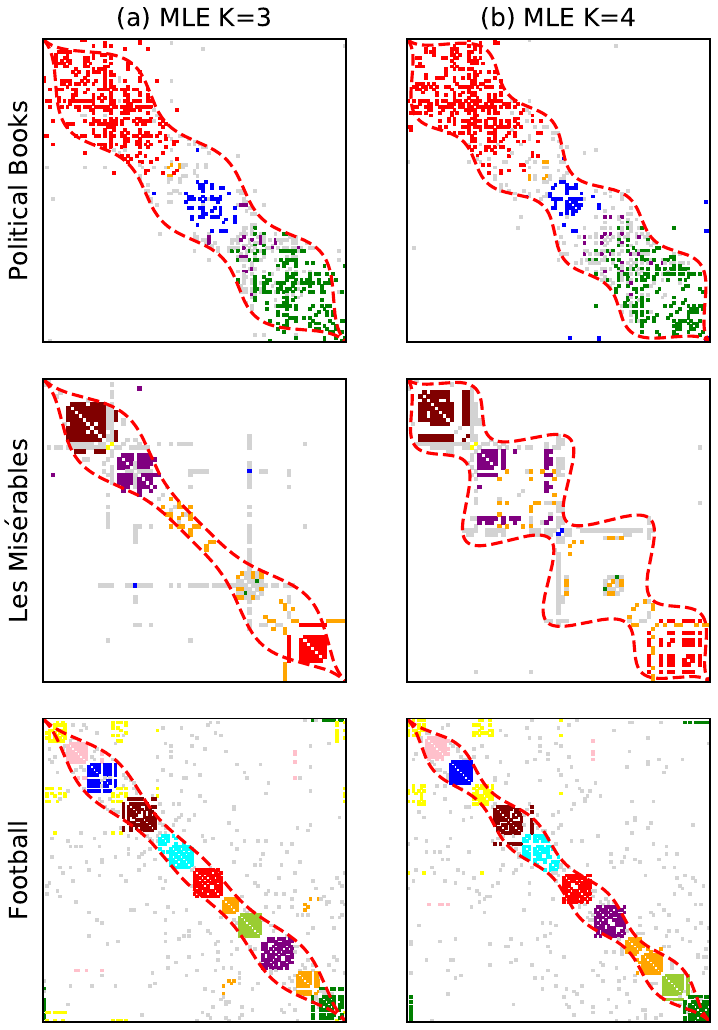}
    \caption{
    Heatmaps of adjacency matrices aligned according to the results of our {\MLE}s of the {\ORGM} with (a) $K=3$ and (b) $K=4$.
    The red dashed line again represents the estimated envelope function. 
    The filling colors of each element represent the same results of community detection as in Fig.~\ref{HeatmapClustering}.
    }
    \label{fig:HeatmapK34}
\end{figure}

\section{\label{sec:appendix} Maximum-likelihood estimate with $K=3$ and $K=4$ on real-world datasets}

In the main text, we showed the result of the {\MLE}s with $K=1$ and $K=2$.
Here, we show the results with of the {\MLE}s with $K=3$ and $K=4$ in Fig.~\ref{fig:HeatmapK34}.
We observe that, except for the result of the Les Mis\'{e}rables network with $K=4$, the inferred vertex sequence is similar to the results with $K=2$, while the envelope functions are more complicated.
In contrast, in the result of the Les Mis\'{e}rables network with $K=4$, the clique-like structure is no longer visible because the log-likelihood function is larger when the elements associated with a hub at the center of the matrix are included in $\Omega_{\textrm{in}}$.
This example illustrates that a larger value of $K$ does not always exhibits a finer community structure and implies the existence of an optimal value of $K$.

\begin{acknowledgments}
This work was supported by JST ACT-X Grant No. JPMJAX21A8 (T.K.), JSPS KAKENHI\ 19H01506, 22H00827 (T.K.), and Quantum Science and Technology Fellowship Program (Q-STEP) (M.O.). 
\end{acknowledgments}

\section*{data availability}
The code of the {\MLE} is available at \cite{github}.

\bibliography{ref.bib}

\clearpage

\end{document}